\documentclass{emulateapj}
 \usepackage{url}

\shorttitle{Magnetar central engine and gravitational wave emission}
\shortauthors{L\"{u} et al}
\slugcomment{}

\begin{document}

\title{Magnetar central engine and possible gravitational wave emission of nearby short
GRB 160821B}
\author{Hou-Jun L\"{u}\altaffilmark{1,2}, Hai-Ming Zhang\altaffilmark{1,2}, Shu-Qing
Zhong\altaffilmark{1,2}, Shu-Jin Hou \altaffilmark{3}, Hui Sun
\altaffilmark{4}, Jared Rice\altaffilmark{5}, and En-Wei Liang\altaffilmark{1,2}}
\altaffiltext{1}{GXU-NAOC Center for Astrophysics and Space
Sciences, Department of Physics, Guangxi University, Nanning
530004, China; lhj@gxu.edu.edu} \altaffiltext{2}{Guangxi Key
Laboratory for Relativistic Astrophysics, Nanning, Guangxi
530004, China} \altaffiltext{3}{College of Physics and
Electronic Engineering, Nanyang Normal University, Nanyang,
Henan 473061, China}\altaffiltext{4}{Department of Astronomy,
School of Physics, Peking University, Beijing 100871,
China}\altaffiltext{5}{Department of Physics and Astronomy,
University of Nevada Las Vegas, NV 89154, USA}

\begin{abstract}
GRB 160821B is a short gamma-ray burst (GRB) at redshift
$z=0.16$, with a duration less than 1 second and without
detection of any ``extended emission'' up to more than 100
seconds in both {\em Swift}/BAT and {\em Fermi}/GBM bands. An
X-ray plateau with a sharp drop 180 seconds after the BAT
trigger was observed with {\em Swift}/XRT. No supernova or
kilo-nova signature was detected. Assuming the central engine
of this SGRB is a recently born supra-massive magnetar, we can
explain the SGRB as jet radiation and its X-ray plateau as the
internal energy dissipation of the pulsar wind as it spins
down. We constrain its surface magnetic field as $B_{\rm p}<
3.12\times 10^{16}$ G and initial spin period as $P_0<
8.5\times 10^{-3}$ seconds. Its equation of state is consistent
with the GM1 model with $M_{\rm TOV} \sim 2.37 M_\odot$ and
ellipticity $\epsilon<0.07$. Its gravitational wave (GW)
radiation may be detectable with the future Einstein Telescope,
but is much weaker than the current detectability limit of
advanced-LIGO. The GW radiation of such an event would be
detectable by advanced-LIGO if it occurred at a distance of 100
Mpc ($z=0.023$).
\end{abstract}


\section{Introduction}
The progenitors of short gamma-ray bursts (SGRBs), which have a
hard spectrum and short duration (Kouveliotou et al. 1993),
remain elusive (Zhang 2011). Several lines of observational
evidence, e.g. low level of star formation (Barthelmy et al.
2005; Berger et al. 2005; Gehrels et al. 2005), a large offset
from the center of the host galaxy (e.g. Fox et al. 2005; Fong
et al. 2010), as well as the non-association of bright
supernovae (SNe) with short GRBs (Berger 2014 and references
therein), suggest that SGRBs may form in compact star mergers,
such as neutron star$-$neutron star mergers (NS$-$NS,
Pacz\'ynski 1986; Eichler et al. 1989), neutron star$-$black
hole mergers (NS$-$BH, Pacz\'ynski 1991), or black hole$-$black
hole mergers (BH$-$BH; Zhang 2016). The coalescence of two
compact stars is also expected to be a strong source of
gravitational wave (GW) radiation, and such systems are the
main targets of the advanced Laser Interferometer
Gravitational-wave Observatory (LIGO)/Virgo detectors. Two GW
events (GW 150914 and GW 151226) and one GW candidate (LVT
151012) were detected with LIGO and are proposed black hole
binary mergers (Abbott et al. 2016a,b). Electromagnetic (EM)
transients associated with gravitational wave bursts (GWBs)
have not been confidently detected, although associations of
weak EM counterparts with these GW events were claimed
(Connaughton et al. 2016). Since the two GW events are believed
to be from BH-BH systems, it is still highly debated whether or
not the merger of a BH$-$BH system can be accompanied by an EM
counterpart (Zhang 2016; Zhang et al 2016; Connaughton et al.
2016; Xiong 2016). Further observations are required to confirm
the existence of BH-BH EM counterparts.

NS-NS mergers as the progenitors of SGRBs have been extensively
studied. Depending on the nascent NS mass ($M_{\rm p}$), two
possible outcomes of the merger are expected. One possibility
is a black hole, which forms when $M_{\rm p}$ is much greater
than the maximum non-rotating mass ($M_{\rm TOV}$, Rosswog et
al. 2003; Rezzolla et al. 2011; Ravi \& Lasky 2014). Another
possibility is a rapidly spinning, strongly magnetized neutron
star (``millisecond magnetar''), in the case where $M_{\rm p}$
is less than $M_{\rm TOV}$ but greater than $M_{\rm max}$ (the
maximum gravitational mass) (Usov 1992; Thompson 1994; Dai \&
Lu 1998a,b; Zhang \& M\'esz\'aros 2001; Metzger et al. 2008,
2011; Bucciantini et al. 2012). The post-merger evolution of
magnetars also depends on the mass lying between $M_{\rm p}$
and $M_{\rm TOV}$. One possible channel is a magnetar in an
equilibrium state which injects energy from the magnetar wind
via loss of rotation energy for $M_{\rm p}\leq M_{\rm TOV}$
(Giacomazzo \& Perna 2013). This well explains the long lasting
energy injection phase observed with the Swift X-Ray Telescope
(XRT; Burrows et al. 2004). Another evolving channel is that of
a supra-massive NS, which may survive if $M_{\rm TOV}<M_{\rm
p}< M_{\rm max}$, when magnetic braking and viscosity compel
the star into uniform rotation. As the period of the magnetar
decreases via rotational energy loss, the maximum gravitational
mass decreases. The magnetar collapses into a black hole when
its centrifugal force cannot balance the gravitational force
(Duez et al. 2006; Ravi \& Lasky 2014). Theoretically, it is
expected that a Poynting-flux dominated outflow is driven by
the injected wind as the magnetar spins down (e.g. Dai \& Lu
1998a; Zhang \& M\'{e}sz\'{a}ros 2001). The observed X-ray
``internal plateau'' (the rapid flux drop off at the end of the
plateau emission with a decay slope $\alpha >
3$)\footnote{Throughout the paper we adopt the convention
$F_\nu \propto t^{-\alpha} \nu^{-\beta}$.} with XRT in a few
long and short GRBs (Troja et al. 2007; Lyons et al. 2010;
Rowlinson et al. 2010, 2013; L\"{u} et al. 2015; Du et al.
2016) may be evidence for this evolution channel. The rapid
decay following the plateau cannot be accommodated in any
external shock model and can be attributed to internal
dissipation of a central engine wind, which is likely a
signature of the collapse of a supra-massive magnetar central
engine into a black hole (Troja et al. 2007; Liang et al. 2007;
Lyons et al. 2010; L\"{u} \& Zhang 2014; L\"{u} et al. 2015).
It has also been proposed that this phenomenon may be
accompanied by a fast radio transient, i.e., fast radio burst
(FRB, Lorimer et al. 2007; Zhang 2014).

In NS-NS merger models, it is predicted that EM signals can not
be avoided after the merger due to the high magnetic field
strength at the NS surface (Metzger \& Berger 2012). In
addition, a NS-NS merger would also lose energy via
gravitational wave quadrupole emission (Fan et al. 2013; Lasky
et al. 2014; Lasky \& Glampedakis 2016). Therefore, hunting for
possible associations of SGRBs with GW events is interesting.
The LIGO team has searched for such associations for many
years, but no events have been reported. Comparing the BH-BH
mergers with the associated GWBs detected with the
advanced-LIGO, the energy lost via GWB in these systems would
be much larger than that expected in NS-NS merger systems
(Corsi \& M\'{e}sz\'{a}ros 2009; Hild et al. 2011; Fan et al.
2013). Therefore, the advanced-LIGO detection rate for NS-NS
mergers should be much lower than that of BH-BH merger systems.

GRB 160821B is a nearby bright SGRB with a redshift of
$z=0.16$.  This paper dedicates analysis of its
multi-wavelength data and constrains the properties of its
central engine as well as its possible GW radiation. We present
our data reduction from {\em Swift} and {\em Fermi}
observations in \S 2. In \S 3, we compare the properties of GRB
160821B with other SGRBs. The derived parameters for a magnetar
central engine and the equation of state of newly-born NSs are
presented in \S 4. In \S5, we present a constraint on the
ellipticity of the NS-NS system and the probability of
detectable gravitational wave radiation. Conclusions are drawn
in \S6 with some additional discussion. Throughout the paper, a
concordance cosmology with parameters $H_0=71~\rm
km~s^{-1}~Mpc^{-1}$, $\Omega_M=0.30$, and
$\Omega_{\Lambda}=0.70$ is adopted.

\section{Data reduction and analysis}

\subsection{{\em Swift} data reduction}
GRB 160821B triggered the Burst Alert Telescope (BAT) at
22:29:13 UT on 2016 August 21 (Siegel et al. 2016). We
developed an IDL script to automatically download the {\em
Swift} BAT data. We use the standard HEASOFT tools (version
6.12) to process the data. We run {\em bateconvert} from the
HEASOFT software release to obtain the energy scale for the BAT
events. The light curves and spectra are extracted by running
{\em batbinevt} (Sakamoto et al. 2007). Then, we calculate the
cumulative distribution of the source counts using the arrival
time of a fraction between 5 and 95 per cent of the total
counts to define $T_{90}$. The time bin size is fixed to 64 ms
in this case due to the short duration. The background is
extracted using two time intervals, one before and one after
the burst. We model the background as Poisson noise, which is
the standard background model for prompt emission. We invoked
Xspec to fit the spectra. For technical details please refer to
Sakamoto et al. (2007). XRT began observing the field 57
seconds after the BAT trigger (Siegel et al. 2016). We made use
of the public data from the {\em Swift} archive\footnote{$\rm
http://www.swift.ac.uk/xrt_curves/00709357$}. The Ultra-Violet
Optical Telescope (UVOT; Roming et al. 2005) observed the field
at $T_0+76$ s, but no optical afterglow was consistent with the
XRT position (Evans et al. 2016). There was also no detection
in the initial UVOT exposures (Xu et al. 2016). Preliminary,
3$\sigma$ upper limits data are obtained by using the UVOT
photometric system for the first finding chart (FC) exposure
(Breeveld et al. 2016). On the other hand, $r$- and $z$-band
afterglow images were obtained by using William Herschel
Telescope on La Palma (Levan et al. 2016). In the spectrum of
the candidate host galaxy several prominent emission lines were
found (H$\beta$, [O$III$] and H$\alpha$), at a redshift of
$z=0.16$. The physical offset of the afterglow from the
candidate host galaxy is approximately 15 kpc (Levan et al.
2016).

\subsection{{\em Fermi} data reduction}
{\em Fermi} Gamma-ray Burst Monitor (GBM) triggered and located
GRB 160821B at 22:29:13.33 UT on 21 August 2016 (Stanbro et al.
2016). GBM has 12 sodium iodide (NaI) and two bismuth germanate
(BGO) scintillation detectors, covering the energy range 8 keV
to 40 MeV (Meegan et al. 2009). We downloaded GBM data of this
GRB from the public science support center at the official
Fermi web site\footnote{http://fermi.gsfc.nasa.gov/ssc/data/}.
Each of the GBM detectors collected the data with three
different types: CTIME, CSPEC, and TTE. Then, we extracted the
light curves and performed spectral analysis based on the
package {\em gtBurst}. By invoking the \texttt{heasoft} command
\texttt{fselect} and the \texttt{ScienceTools} command
\texttt{gtbin}, we extracted light curves with a time-bin of 64
ms in a user-specified energy band from the GBM. We clicked
``Tasks $\rightarrow$ Make spectra for XSPEC'' in $gtBurst$ to
extract the source spectrum of the GBM data. The background
spectra are extracted from the time intervals before and after
the prompt emission phase, modeled with a polynomial function,
and the source spectrum is extracted by applying the background
model to the prompt emission phase. This GRB occurred
right at the edge of the Large Area Telescope (LAT)
field-of-view (Atwood et al. 2009), about 61 degrees from
boresight. So we do not expect to detect the high-energy
signal if it exists.

\subsection{{\em Swift} and {\em Fermi} data analysis}
As shown in Fig.1, the BAT light curve shows a single short
peak with duration $T_{90}=0.48\pm0.07$ seconds, and there is
no evidence of extended emission detected in the BAT energy
range up to 100 s after the BAT trigger ($T_0$). The
time-integrated BAT spectrum can be fit by a single power law
with photon index $\Gamma_{\gamma}=1.98\pm0.11$. The BAT band
(15$-$150 keV) peak flux is $(1.7\pm0.2)~\rm photons~cm^{-2}~s
^{-1}$, and the total fluence is $(1.1\pm0.1)\times10^{-7} ~\rm
erg ~cm^{-2}$ (Palmer et al. 2016). The GBM light curve of GRB
160821B is also shown in Figure 1 with a 64 ms time-bin. The
profile of light curve is similar with BAT data, has a bright
peak with duration $T_{90}\sim 1.2$ s in 8$-$1000 keV range.
The GBM spectra can be fit by a power law function due to lack
high energy photons\footnote{Stanbro et al. (2016) find that
the spectrum can be fit by a power law function with an
exponential high-energy cutoff. The power law index is
$\Gamma_{\rm GBM}=-1.31\pm0.6$, and the cutoff energy $E_{\rm
p}=84\pm 19~\rm keV$.}. Here, we joint fit the time-averaged
spectra of {\em Fermi}/GBM+{\em Swift}/BAT with a power-law
model, and found $\Gamma_{\rm BAT+GBM}=1.88\pm0.12$ with
$\chi^{2}=0.84$ (Fig. 2). The total fluence in 8-10000 keV is
$(2.52\pm0.19)\times10^{-6} ~\rm erg ~cm^{-2}$. The isotropic
energy $E_{\rm \gamma, iso}=(2.1\pm0.2)\times 10^{50}~\rm erg$.

Norris et al (2000) discovered an anti-correlation between GRB
peak luminosity and the delay time ($t_{\rm lag}$) in different
energy bands, meaning softer photons usually arrive later than
hard photons. This spectral lag is always significant in
long-duration GRBs (Norris et al. 2000; Gehrels et al. 2006;
Liang et al. 2006), but is typically negligible in
short-duration GRBs (Norris \& Bonnell 2006; Zhang et al.
2009). We extracted 4 ms binned light curves in the following
three BAT energy bands: 15-25 keV, 25-50 keV, 50-100
keV\footnote{The signal in 100-150 keV is too weak to be
extracted, so we do not consider the emission in this energy
band.}. Then, we used the cross-correlation function method
(CCF; Norris et al. 2000; Ukwatta et al. 2010) to calculate the
lags between 25-50 keV and 50-100 keV light curves. In
order to address the questions of whether the short GRB 160821B
is consistent with typical Type I and other short-hard GRBs
with in the spectral lag distribution. Figure 4 shown the peak
luminosity as function of spectral lag for typical Type II,
Type I, other short-hard, and GRB 160821B. Here, Type II GRBs
are corresponding to confirmed supernova (SN) association, or
have a high specific star formation rate (SSFR) and do not have
a large offset from the galaxy center. On the contrary, Type I
GRBs are occured in elliptical or early type host galaxy
without SN signature, or has a relatively low local SSFR and
large offset from the host galaxy center. For the other
short-hard GRBs, do not satisfy neither of the two criteria of
the Type I sample, and do not have their host galaxy
identified, but with a short duration, hard spectral (Zhang et
al. 2009). Type II GRBs sample are from Norris \& Bonnell
(2006), and give the best power-law model fitting with
$2\sigma$ region of the fitting. Type I and other short-hard
GRBs are collected from Zhang et al (2009). For GRB 160821B, we
found $t_{\rm lag}=(10\pm6)$ ms, which is consistent with
result of Palmer et al (2016). It is deviated from $2\sigma$
region of the Type II GRBs fitting, but both peak luminosity
and lag value of this case is comparable with Type I GRBs.

The initial X-ray light curve is best fit by a broken power
law, which reads
\begin{equation}
F=F_0 \left[\left(\frac{t}{t_{\rm b}}\right)^{\omega {\alpha_1}}+\left(\frac{t}{t_{\rm
b}}\right)^{\omega
{\alpha_2}}\right]^{-1/\omega},
\end{equation}
where $\omega$ describes the sharpness of the break and is
taken to be 3 in this analysis (Liang et al. 2007). There is an
initial decay slope $\alpha_1=0.21\pm0.14$, followed by a
steeper decay of $\alpha_2=4.52\pm0.45$ with a break time
$t_{b}=180\pm46$ seconds after the BAT trigger. No significant
X-ray flare was detected during the observational time. The
X-ray spectrum in the 0.3$-$10 keV energy band is best fit by
an absorbed power law with $\Gamma_X=1.95^{+0.21}_{-0.08}$ and
column density $N_{\rm H} = (7.5\pm2.1)\times10^{20} ~\rm
cm^{-2}$. The X-ray light curve along with the $\Gamma_{\rm X}$
evolution is shown in Fig.3. About 1000 seconds after the BAT
trigger, another component emerged which is likely a normal
decay and post-jet break. We used a broken power law to fit
this component and found $\alpha_3 \sim 0.45$, $\alpha_4 \sim
3.5$ with break time around 35000 seconds. We follow the method
discussed in Zhang et al. (2007) to calculate $E_{\rm K,iso}$,
which is almost constant during the normal decay phase in X-ray
afterglow. We assume that it is in the $\nu > \rm max(\nu_{\rm
m}, \nu_{\rm c})$ region, where the afterglow flux expression
does not depend on the medium density. In our calculations, the
microphysics parameters of the shock are assigned standard
values derived from observations, i.e. $\varepsilon_e=0.01$ and
$\varepsilon_B=0.001$, and thus $E_{\rm K, iso}\sim 8\times
10^{52} ~\rm erg$. If the later break is assumed to be a
post-jet break\footnote{The error bar of the last X-ray data
point is large and thus it is difficult to identify where the
jet break occurs. However, it is possible to provide a lower
limit to the jet opening angle if we assume that it is a jet
break.}, one can estimate the jet opening angle
$\theta_j\sim0.063~\rm rad\sim 3.6$ degrees with medium density
$n=0.1~\rm cm^{-3}$.

The XRT light curve of short GRB 160821B that has a short
plateau emission following an abrupt decay is unusual, but not
odd. This temporal behavior is similar to short GRB 090515
(Rowlinson et al. 2010). In Figure 5(a), we collect all of the
short GRB light curves without extended emission to compare
with the X-ray emission of GRB 160821B. We found that most
short GRBs appeared to have a steeper decay around several
hundred seconds. Particularly, the X-ray emission behavior of
GRB 160821B is similar to GRB 090515, which is the first short
GRB claimed to have a magnetar central engine origin (Rowlinson
et al. 2010). Also, the plateau flux of GRB 160821B is the
highest compared to other short GRBs. In Figure 5(b), we show
the fluence in the BAT band (15$-$150 keV) and flux in the XRT
band (0.3$-$10 keV) at $T_0+100$ s for all the short GRBs in
the {\em Swift} sample. GRB 160821B is shown with a filled
circle. As expected, the higher fluence GRBs tend to have
higher flux X-ray afterglows.

\section{The central engine of GRB 160821B}
The abrupt decay following the bright X-ray plateau observed in
GRB 160821B is difficult to explain by invoking the external
shock model of the black hole central engine. It must invoke
the contributions from internal dissipation of a central
engine. In this section, we propose the use of the millisecond
magnetar central engine model to explain the abrupt decay
behavior in the X-ray afterglow emission, and constrain the
parameters of the magnetar.

\subsection{Magnetar central engine}
According to Zhang \& M\'esz\'aros (2001), the characteristic
spin down luminosity $L_0$ and time scale $\tau$ are written:
\begin{equation}
L_0 = 1.0 \times 10^{49}~{\rm erg~s^{-1}} (B_{\rm p,15}^2 P_{0,-3}^{-4} R_6^6),
\label{L0}
\end{equation}
\begin{equation}
\tau = 2.05 \times 10^3~{\rm s}~ (I_{45} B_{\rm p,15}^{-2} P_{0,-3}^2 R_6^{-6}),
\label{tau}
\end{equation}
where $I$ is the moment of inertia of a typical NS with mass
$M_{\rm NS}=1.4 M_{\odot}$, $P_0$ is the initial spin period,
$B_{\rm p}$ is the magnetic field strength, $R$ is the radius
of the NS, and the convention $Q=10^x Q_x$ is adopted in cgs
units for all other parameters throughout the paper. The
spin-down time scale can be generally identified as the lower
limit of the observed break time, i.e.
\begin{equation}
\tau > t_{b}/(1+z),
\label{tau = tb}
\end{equation}
where $t_{\rm b}$ is the break time after the internal plateau
found using a broken power-law function fitting. A redshift
$z=0.16$ is adopted. The bolometric luminosity at the break
time $t_{\rm b}$ is:
\begin{eqnarray}
L_b = 4\pi D_L^2 F_b \cdot k,
\end{eqnarray}
where $F_b=(1.6\pm0.82)\times 10^{-9}~\rm erg~cm^{-2} s^{-1}$
is the X-ray flux at $t_{b}$, $D_L^2$ is luminosity distance,
and $k$ is $k$-correction factor. The characteristic spin-down
luminosity is essentially the plateau luminosity, which may be
estimated as
\begin{eqnarray}
L_0 \simeq L_{\rm b}=(1.8\pm0.6)\times 10^{48}~\rm erg~s^{-1}.
\label{L0b}
\end{eqnarray}

Based on Equation (\ref{L0}) and Equation (\ref{tau}), one
can derive the magnetar parameters $B_{\rm p}$ and $P_0$:
\begin{eqnarray}
B_{\rm p,15} = 2.05(I_{45} R_6^{-3} L_{0,49}^{-1/2} \tau_{3}^{-1})~\rm G,
\label{Bp}
\end{eqnarray}
\begin{eqnarray}
P_{0,-3} = 1.42(I_{45}^{1/2} L_{0,49}^{-1/2} \tau_{3}^{-1/2})~\rm s.
\label{P0}
\end{eqnarray}
Using the lower limit of $\tau$ we derive upper limits for
$P_0$ and $B_{\rm p}$, which are respectively $P_0 < 8.5 \times
10^{-3}$ s and $B_{\rm p}< 3.12 \times 10^{16}$ G. Figure 6a
shows the $B_{\rm p}-P_0$ diagram for GRB 160821B, and compares
other short GRBs.

\subsection{Equation of state of NS}
Another relevant timescale is the collapse time of a
supra-massive magnetar, $t_{\rm col}$. The post-internal
plateau decay slope $\alpha_2$ is steeper than 2, which is the
standard spin down luminosity evolution with time (Zhang \&
M\'esz\'aros, 2001). The break time is therefore defined by the
collapse time $t_{\rm col}$, and one can write
\begin{equation}
\tau \simeq t_{\rm b}.
\label{tcol = tb}
\end{equation}
The maximum gravitational mass ($M_{\rm max}$) depends on spin
period which increases with time. Using the same method
description in Lasky et al. (2014) and L\"{u} et al. (2015), we
can write down $t_{\rm col}$ as a function of $M_{\rm p}$
\begin{eqnarray}
t_{\rm col} &=& \frac{3c^{3}I}{4\pi^{2}B_{\rm p}^{2}R^{6}}[(\frac{M_{\rm p}-M_{\rm
TOV}}{\hat{\alpha} M_{\rm
TOV}})^{2/\hat{\beta}}-P_{0}^{2}]\nonumber \\
&=&\frac{\tau}{P_{\rm 0}^{2}}[(\frac{M_{\rm p}-M_{\rm TOV}}{\hat{\alpha} M_{\rm
TOV}})^{2/\hat{\beta}}-P_{0}^{2}].
\label{tcol}
\end{eqnarray}
where $\hat{\alpha}$, $\hat{\beta}$, and $M_{\rm TOV}$ are
dependent on the equation of state. Therefore we can use
$t_{\rm col}$ to constrain the NS equation of state (EOS).
Here, we only consider five EOS (SLy, APR, GM1, AB-N, and AB-L)
for the given proto-magnetar mass distribution derived from the
total mass distribution of Galactic NS$-$NS binary systems
(Fig.6b). (1) SLy: is effective nuclear interaction by neutron
rich matter with $M_{\rm TOV}=2.05 M_{\odot}$ and $R=9.97 \rm
km$. (2) APR: assume that the inner material is included both
dense nucleon admixture of quark materr, with $M_{\rm TOV}=2.20
M_{\odot}$ and $R=10.00 \rm km$. (3) GM1: by relating scalar
and vector couplings of the hyperons for saturated nuclear
matter with $M_{\rm TOV}=2.37 M_{\odot}$ and $R=12.05 \rm km$.
(4) AB-N: neutrons nuclear attraction due to pion exchange
tensor with $M_{\rm TOV}=2.67 M_{\odot}$ and $R=12.90 \rm km$.
(5) AB-L: neutrons nuclear attraction due to scalar exchange
with $M_{\rm TOV}=2.71 M_{\odot}$ and $R=13.70 \rm km$ (Lasky
et al. 2014). Our results show that the GM1 model gives an
$M_{\rm p}$ band falling within the $2\sigma$ region of the
protomagnetar mass distribution, such that the GM1 EOS is the
best candidate for a non-rotating NS with maximum mass $M_{\rm
TOV}=2.37 M_{\odot}$.

\subsection{The energy budget of magnetar}
One of the most important necessary conditions of a magnetar
central engine candidate for GRBs is that the sum of the prompt
emission energy ($E_{\rm \gamma, iso}$), internal plateau
energy ($E_{\rm X, iso}$), and kinetic energy ($E_{\rm K,
iso}$) after jet correction should be less than the total
rotation energy (energy budget of magnetar) if we assume the
magnetar wind is isotropic. The total rotation energy of the
millisecond magnetar is
\begin{equation}
E_{\rm rot} = \frac{1}{2} I \Omega_{0}^{2}
\simeq 2 \times 10^{52}~{\rm erg}~
M_{1.4} R_6^2 P_{0,-3}^{-2},
\label{Erot}
\end{equation}
where $\Omega_0 = 2\pi/P_0$ is the initial angular frequency of
the neutron star. The total energy of the prompt emission is
$E_{\rm \gamma, iso}=(2.1\pm0.2)\times 10^{50}~\rm erg$ within
the energy range 8-1000 keV. The X-ray internal plateau energy
can be roughly estimated using the break time and break
luminosity (L\"u \& Zhang 2014), i.e.
\begin{eqnarray}
E_{\rm X, iso} &\simeq& L_b \cdot \frac{t_b}{1+z} \nonumber \\
&\simeq& (2.79\pm 0.9)\times 10^{49}~\rm erg.
\end{eqnarray}
To estimate the kinetic energy $E_{\rm K, iso}$, which is used
in the standard forward afterglow model, one has $E_{\rm
K, iso}\sim 8\times 10^{52} ~\rm erg$ (see section 2.3). Therefore
$E_{\rm rot}\gg \frac{1}{2}\theta^{2}_{\rm
j}(E_{\rm \gamma, iso}+E_{\rm X, iso}+E_{\rm K, iso})$, which
satisfies the magnetar central engine energy budget requirement.

\section{Gravitational wave constraints}
\subsection{Ellipticity constraints of newly-born NS}
The coalescence of double neutron stars is believed to be one
of the most likely sources for powering gravitational wave
radiation with associated EM signals. These events have
promising detectability prospects with current and future
gravitational wave detectors like advanced LIGO/Virgo (Zhang
2013; Gao et al. 2013; Yu et al. 2013; Fan et al. 2013). If
indeed a magnetar drove GRB 160821B, why was the total rotation
energy of the magnetar much larger than the sum of the prompt
emission energy, internal energy and kinetic energy? Several
possible reasons may be used in interpreting the gap in the
energy compared to the energy budget. One is that the
efficiency is as low as $\sim 0.01$, such low efficiency may
disfavor the magnetic energy dissipation process (Fan et al.
2013). Another possibility is the missing energy must have been
carried away by non-electromagnetic gravitational wave
radiation (Fan et al. 2013; Lasky \& Glampedakis 2016; Ho
2016), or carried to the black hole before spin down.

Following Fan et al. (2013) and Lasky \& Glampedakis (2016), a
magnetar loses rotational energy through two channels: magnetic
dipole torques ($L_{\rm m}$) gravitational wave radiation ($L_{\rm w}$)
\begin{eqnarray}
-dE_{\rm rot}/dt &=& L_{\rm m} + L_{\rm w} \nonumber \\
&=& \frac{B^2_{\rm
p}R^{6}\Omega^{4}}{6c^{3}}+\frac{32GI^{2}\epsilon^{2}\Omega^{6}}{5c^{5}},
\end{eqnarray}
where $\epsilon=2(I_{\rm xx}-I_{\rm yy})/(I_{\rm xx}+I_{\rm
yy})$ is the ellipticity in terms of the principal moments of
inertia, assuming the magnetar has a triaxial shape. Following
the method of Lasky \& Glampedakis (2016), GW radiation can be
more efficient than magnetic dipole radiation because of its
stronger dependence on the neutron star spin rate $\Omega$,
i.e. $\Omega^{6}$ and $\Omega^{4}$ respectively. The upper
limit on the ellipticity ($\epsilon$) can be expressed simply
with a dependence on observed plateau luminosity and break time
(Lasky \& Glampedakis 2016),
\begin{eqnarray}
&\epsilon_{\rm obs}& \leq \biggl(\frac{15c^{5}\eta I}{512G L^{2}_{0} t^{3}_{\rm b}}\biggr)^{1/2}
\nonumber \\
&=& \!0.33\eta\biggl(\frac{I}{10^{45}~\rm g\,cm^{2}}\biggr)^{\!1/2}\!\biggl(\frac{L_{0}}{10^{49}~\rm
erg\,s^{-1}}\biggr)^{\!-1}\!\biggl(\frac{t_{\rm b}}{100~\rm
s}\biggr)^{\!-3/2}.\nonumber
\\
\end{eqnarray}
Using the typical NS mass and radius, $\eta=0.1$, $L_{0}\sim
5.38\times 10^{47}~\rm erg~s^{-1}$, and $t_{\rm b}\sim 180$ s,
one has $\epsilon_{\rm obs}< 0.07$.

\subsection{Detection probability of gravitational wave}
If most of the rotation energy is released via gravitational wave
radiation with a frequency $f$, the gravitational wave strain
for a rotating neutron star at distance $D_{\rm L}$ can be
expressed as
\begin{eqnarray}
h(t)=\frac{4G I \epsilon}{D_{\rm L}c^{4}} \Omega(t)^{2}.
\end{eqnarray}
The noise power spectral density of the detector, $S_{h}(f)$,
and the stationary phase approximation implies
$\tilde{h}(f)^{2}=h(t)^{2}|dt/df|$, where $\tilde{h}(f)$ is the
Fourier transform of $h(t)$. Following method of Lasky \&
Glampedakis (2016), $\tilde{h}(f)$ can be expressed as
\begin{eqnarray}
\tilde{h}(f) &=& \frac{1}{D_{\rm L}}\sqrt{\frac{5GI}{2c^{3}f}} \nonumber \\
&\approx& 2.6\times 10^{-25} \biggl(\frac{I}{10^{45}~\rm g\,cm^{2}}\frac{1~\rm kHz}{f}\biggr)^{\!1/2}\biggl(\frac{D_{\rm L}}{100~\rm
Mpc}\biggr)^{\!-1}. \nonumber \\
\end{eqnarray}
So $\tilde{h}(f)$ is independent of the neutron star ellipticity,
but depends on the angular frequency evolution with time.
The characteristic amplitude $h_{\rm c}=f h(t)\sqrt{dt/df}=f
\tilde{h}(f)$ (Corsi \& M\'{e}sz\'{a}ros 2009; Hild et al.
2011) is
\begin{eqnarray}
h_{\rm c} &=& \frac{f}{D_{\rm L}}\sqrt{\frac{5GI}{2c^{3}f}} \nonumber \\
&\approx& 8.22\times 10^{-24} \biggl(\frac{I}{10^{45}~\rm g\,cm^{2}}\frac{f}{1~\rm kHz}\biggr)^{\!1/2}\biggl(\frac{D_{\rm
L}}{100~\rm Mpc}\biggr)^{\!-1}~\rm.
\nonumber
\\
\end{eqnarray}
For GRB 160821B, its redshift $z=0.16$ corresponds to $D_{\rm
L}=765~\rm Mpc$. Using this and $f=1000~\rm Hz$, one can
estimate the maximum value of the strain $h_{\rm c}$, which is
less than $1.1\times 10^{-24}$. In Fig.7, we plot the
gravitational wave strain sensitivity for advanced-LIGO and
Einstein Telescope (ET), from Figure 3 of Lasky \& Glampedakis
(2016). It is clear that the strain of GRB 160821B is below the
initial LIGO or advanced-LIGO noise curve. However, it is
comparable to the proposed detectability limit of ET and such a
signal may be detected by ET in the future.

On the other hand, keeping the total energy constant and moving
the event to a lower redshift allows one to estimate the
minimum detectability distance of such an event. The
gravitational wave strain amplitude will be stronger if the
event occurs at a lower redshift. One can estimate the
cosmological distances the gravitational wave signal can be
detected by current advanced-LIGO. We simulate this source at
different distances and calculate the GW strain amplitude. Then
we compare that value with the current sensitivity of
advanced-LIGO. We find that this GW signal could be detected if
shifted to about $100~\rm Mpc$, which corresponds to redshift
$z\sim 0.023$ (Fig. 7).

\section{Conclusions and Discussion}
GRB 160821B is a short gamma-ray burst (GRB) of duration less
than 1 second, at redshift $z=0.16$, observed by {\em Swift}
and {\em Fermi}. We presented a broadband analysis of its
prompt and afterglow emission and found that there is no
evidence to for any ``extended emission'' up to more than 100
seconds in {\em Swift}/BAT and {\em
Fermi}/GBM\footnote{Presence or absence of extended emission of
short GRB may be related to different physics process.}. More
interestingly, the X-ray plateau was followed by an extremely
steep decay as observed by {\em Swift}/XRT but which is not
unique in the {\em Swift} era, i.e. it is similar to GRB 090515
(Rowlinson et al. 2010), which was the first short GRB with
such behavior. This behavior is very difficult to explain with
the standard external shock model of black hole central engine,
but could be consistent with the prediction of a magnetar
central engine. It is likely that it formed into supra-massive
NS initially and collapsed into black hole after several
hundred seconds. This event is thus one important probe for
studying the physical properties of the central engine and
progenitor of GRBs.

Our analysis shows the initial short plateau emission in its
X-ray lightcurve, which is consistent with energy injection
from the magnetar wind of a supra-massive magnetar losing
rotation energy, and followed by a steeper decay due to the
magnetar collapsing to a black hole. The derived magnetar
surface magnetic field $B_{\rm p}$ and the initial spin period
$P_0$ fall into a reasonable range, i.e. $B_{\rm p}<3.12\times
10^{16}$ and G$P_0 < 8.5\times 10^{-3}$ s. Using the collapse
time to constrain the equation of state of the neutron star
shows consistency with the GM1 model with $M_{\rm TOV} \sim
2.37 M_\odot$. The total isotropic-equivalent electromagnetic
energy ($\gamma$-ray energy, internal plateau energy, and
kinetic energy) is much less than the energy budget of the
magnetar (a few $\times 10^{52}~\rm erg$), suggesting that the
missing energy of the supra-massive magnetar may be radiated
via gravitational waves, or carried into the black hole before
spin down. If it is indeed that the energy dissipated via
gravitational waves, one can constrain the ellipticity of the
NS to $\epsilon<0.07$. Also, the upper limit of the
gravitational wave strain can be estimated as $h_{\rm c}
\approx 1.1\times 10^{-24}$ at $f=1000~\rm Hz$, which is below
the advanced-LIGO noise curve, but may be detectable by
Einstein Telescope in the future. If we shift this source to
$\sim 100~\rm Mpc$ cosmological distance ($z\sim 0.023$), then
the gravitational wave signal could be detected by the current
advanced-LIGO.

The event rate density of SGRBs depends on the minimum
luminosity threshold. Given the detectability horizon of
advanced-LIGO, i.e. a distance of 100 $\rm Mpc$, all the
observed SGRBs are above the BAT sensitivity. Following the
method of Sun et al. (2015), if we consider the minimum
isotropic luminosity of the observed SGRBs, which gives an
event rate of $4.2^{+1.3}_{-1.0}~\rm Gpc^{-3}~yr^{-1}$ above
$7\times10^{49}~\rm erg~s^{-1}$ and varies by a factor less
than two for different delay timescale models, we estimate
there are 2 SGRBs every one hundred years within 100 $\rm Mpc$.
This is quite small and consistent with the non-detection of
any SGRBs accompanying detected GW events. It is also possible
that there may be low-luminosity SGRBs extending to a
luminosity of $10^{47}~\rm erg~s^{-1}$, which is the detection
limit for {\em Swift}/BAT for SGRBs at 100 $\rm Mpc$. The
estimated event rate density above this luminosity threshold is
much larger than that of the $7\times10^{49}~\rm erg~s^{-1}$
luminosity threshold. In this case, one may expect one low
luminosity SGRB every two years. However, this is quite
speculative as we have not seen any low luminosity SGRBs yet.
We have already had a few such cases for LGRBs. Both of the two
cases can be tested with future detections. For current
circumstances, the first scenario is preferred.

On the other hand for NS-NS mergers a more isotropic,
sub-relativistic outflow could be ejected during a neutron-rich
merger which can synthesize heavier radioactive elements via
r-process. A thermal UV-optical transient may be powered by
radioactive decay except the short GRB and its X-ray afterglow
(Li \& Paczynski 1998; Rezzolla et al. 2011; Yu et al. 2013).
However, if the post-merger product is a supra-massive NS
supported by rigid rotation, e.g. GRB 160821B, the spin-down
magnetic dipole radiation of the NS remnant provides an
additional energy source to the ejecta. This optical transient
(Li-Paczynski-nova, macro-nova, kilo-nova, merger-nova,
r-process) emission component would be significantly enhanced
since it is heated by the magnetar wind and could easily exceed
the r-process power (Li \& Paczynski 1998; Tanvir et al. 2013;
Berger et al. 2013; Yu et al. 2013; Yang et al. 2015; Gao,
Zhang \& L\"{u} 2016; Gao et al. 2016). From the theoretical
point of view, it is expected that the optical or near-infrared
bump is detected at late time or an excess of flux would be
visible in the spectral energy distribution. One can use the
properties of observed merger-nova to constrain the parameters
of the central engine. However, due to lack of optical
observations, catching the possible merger-nova is expected by
following up with an optical instrument in the future, i.e.
Hubble Space Telescope (HST).

\acknowledgments

We acknowledge the use of the public data from the {\em Swift},
{\em Fermi} data archive, and the UK {\em Swift} Science Data
Center. We also thank the anonymous referees for helpful
comments. This work is supported by the National Basic Research
Program (973 Programme) of China 2014CB845800, the National
Natural Science Foundation of China (Grant No.11603006,
11533003, 11503011), the One-Hundred-Talents Program of Guangxi
colleges, Guangxi Science Foundation (grant No.
2016GXNSFCB380005, 2013GXNSFFA019001), Scientific Research
Foundation of GuangXi University (Grant No XGZ150299).

\begin{figure}
\includegraphics[angle=0,scale=0.6]{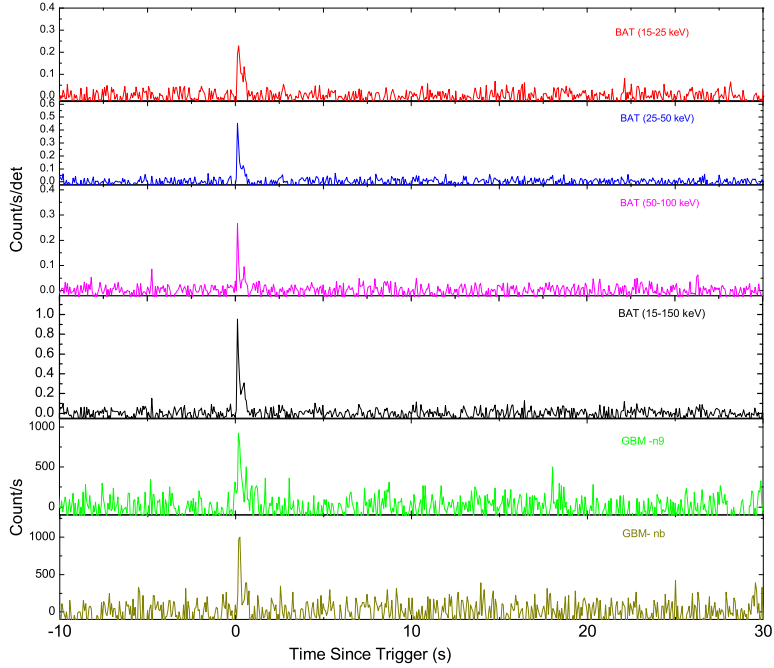}
\hfill\caption{The {\em Swift}/BAT and {\em Fermi}/GBM lightcurve of GRB 160821B
in different energy bands with 64 ms time bin.}
\end{figure}

\begin{figure}
\includegraphics[angle=-90,scale=0.5]{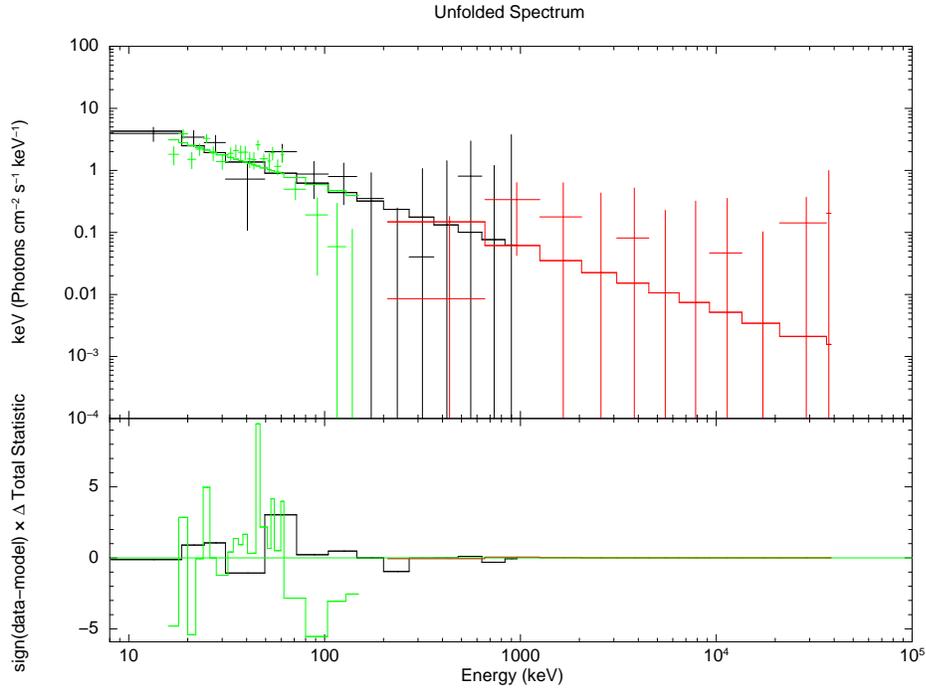}
\hfill\caption{Joint fit of the time-averaged spectra of BAT (green points)
and GBM (red and black points) data with a power-law model.}
\end{figure}


\begin{figure}
\includegraphics[angle=0,scale=0.6]{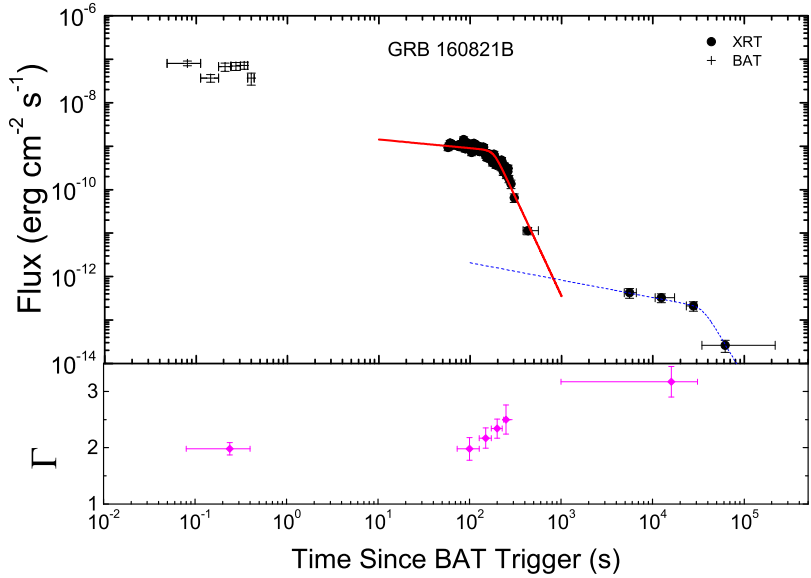}
\hfill\caption{The {\em Swift}/XRT light curve of GRB 160821B (black points). The
{\bf lower} plot shows the photon index evolution. The red solid line and blue dashed line show the
broken power-law fit to the light curve.}
\end{figure}

\begin{figure}
\includegraphics[angle=0,scale=0.6]{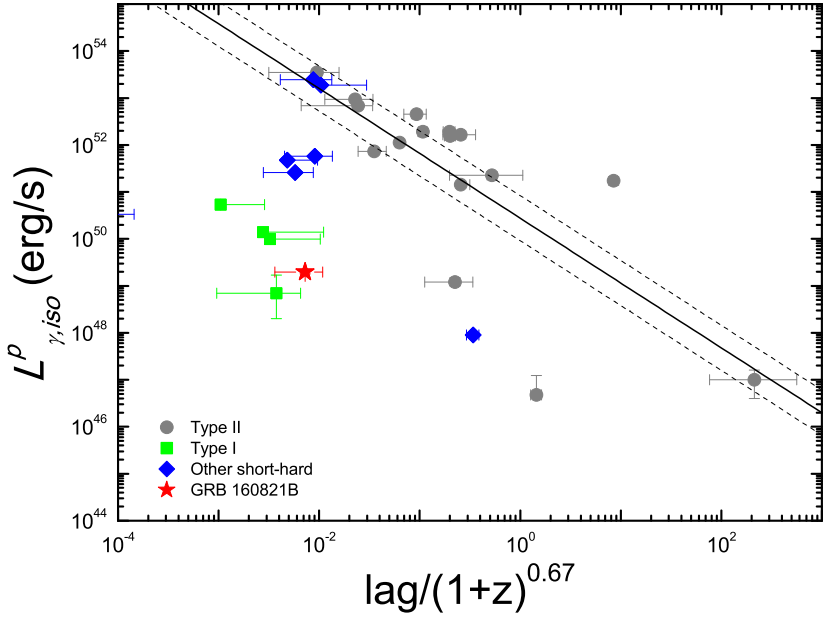}
\hfill\caption{Luminosity-spectral lag diagram. The red star indicates GRB 160821B.
Grey dots, green squares, and blue diamonds are indicated Type II, Type I, and other-short hard GRBs, respectively.
The solid black line and two dash lines are represented the best linear fit to type II GRBs and $2\sigma$ region.}
\end{figure}


\begin{figure}
\includegraphics[angle=0,scale=0.4]{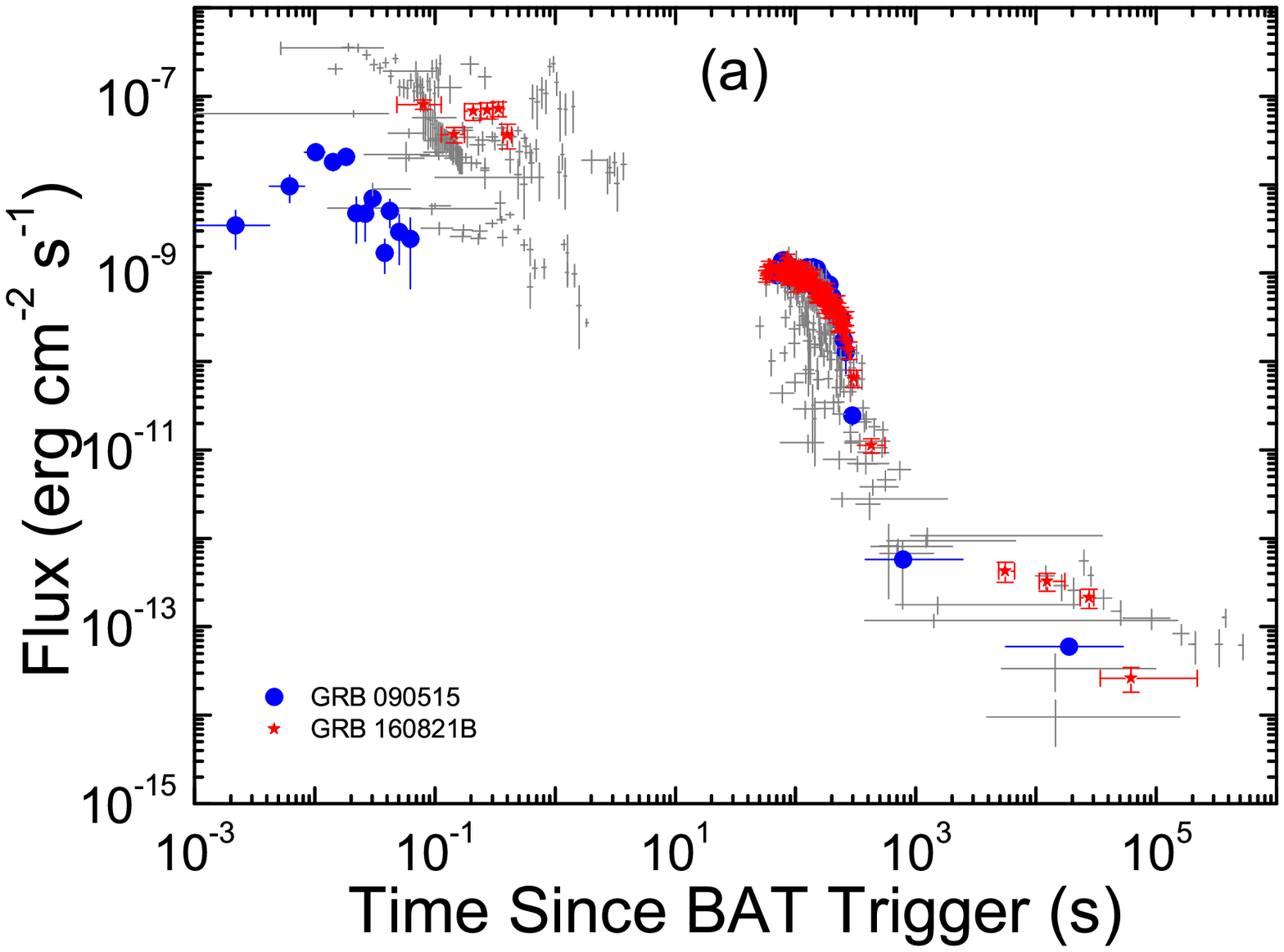}
\includegraphics[angle=0,scale=0.4]{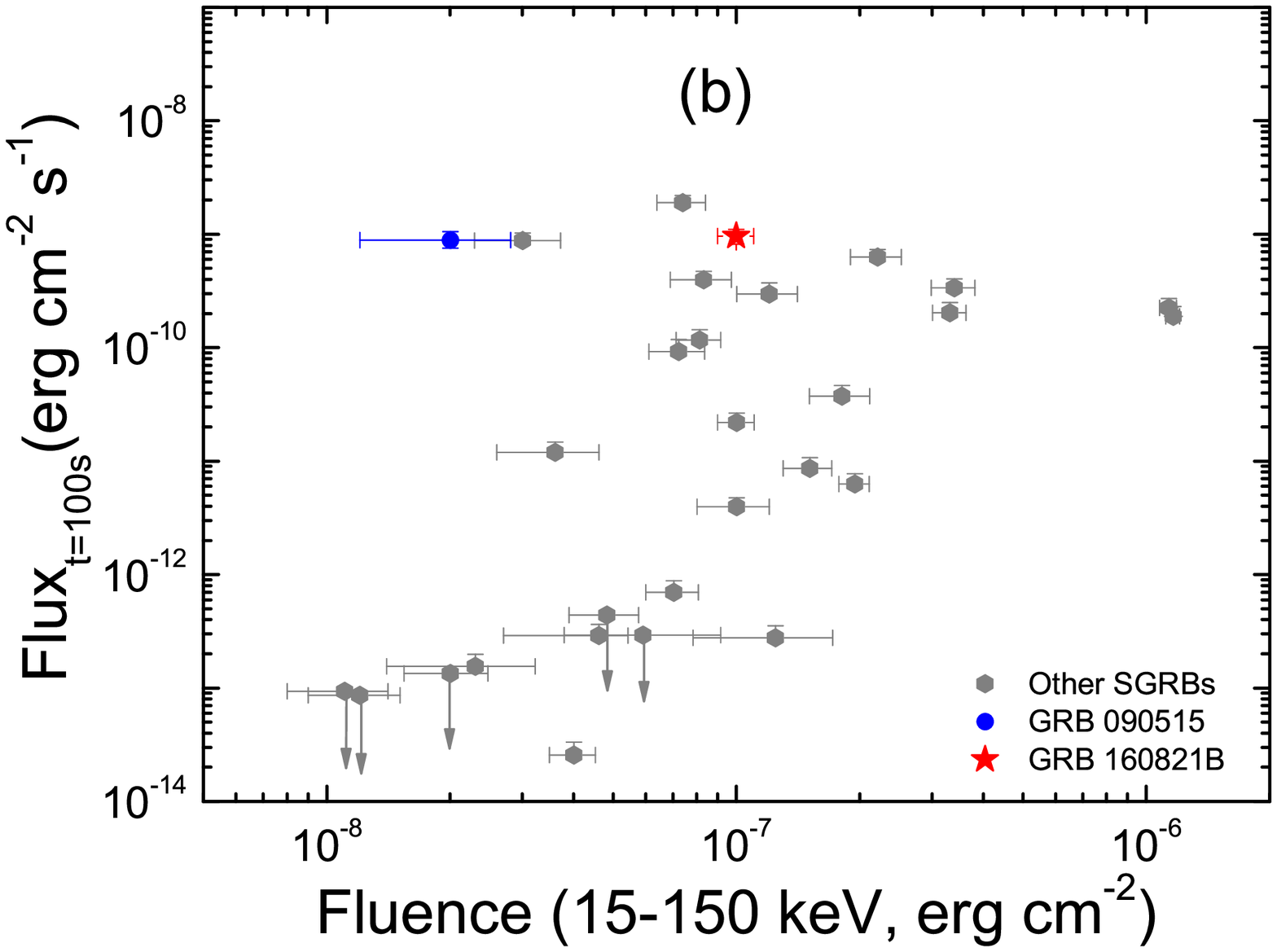}
\hfill\caption{(a): Comparing the X-ray light curve (0.3$-$10 keV) of GRB 160821B
with other short GRBs without extended emission.
(b): The fluence in BAT energy band (15$-$150
keV) versus the flux in XRT band (0.3$-$10 keV) for all Swift SGRBs which were
observed 100s after the trigger
time. The red star marks the location of GRB 160821B.}
\end{figure}


\begin{figure}
\includegraphics[angle=0,scale=0.4]{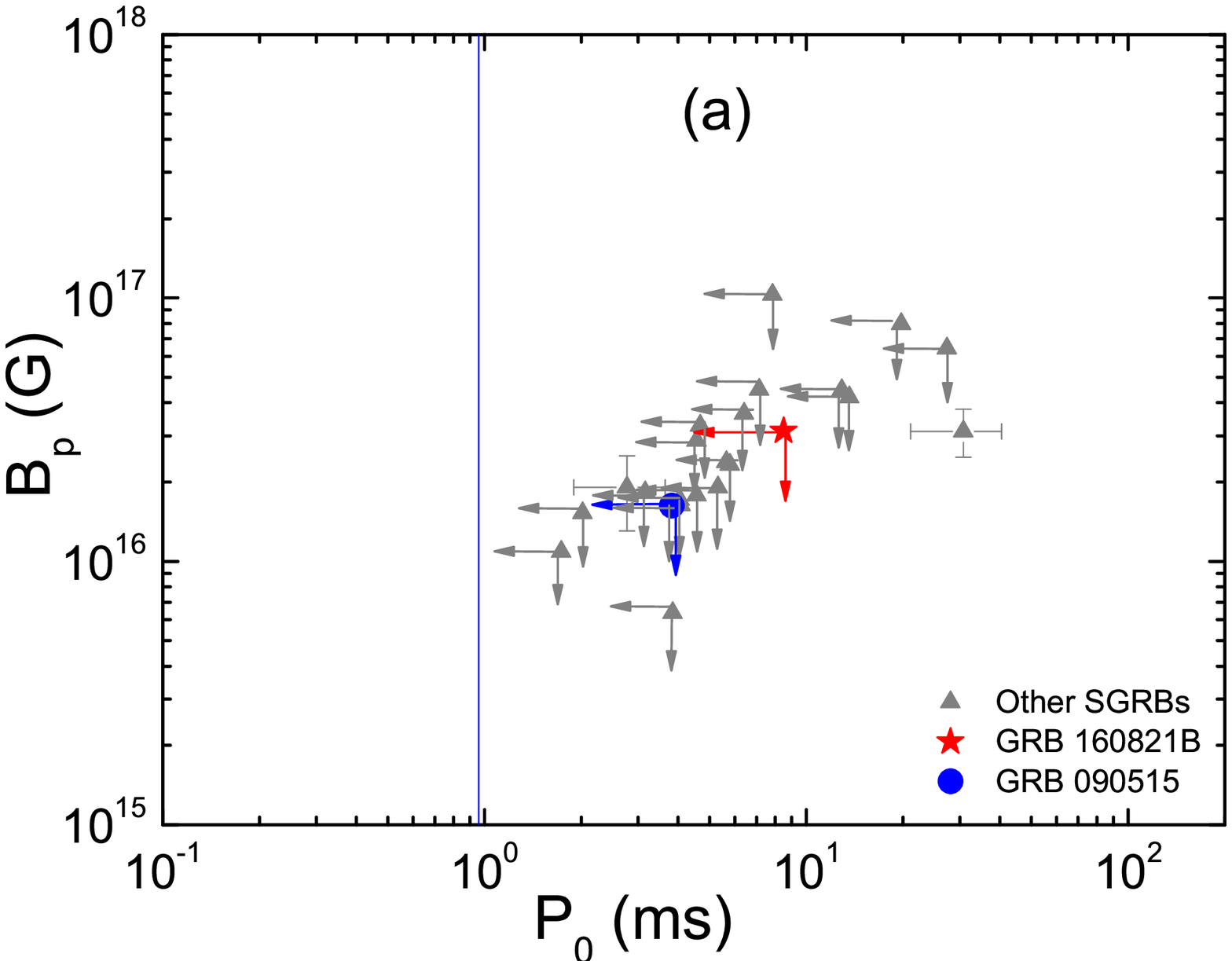}
\includegraphics[angle=0,scale=0.4]{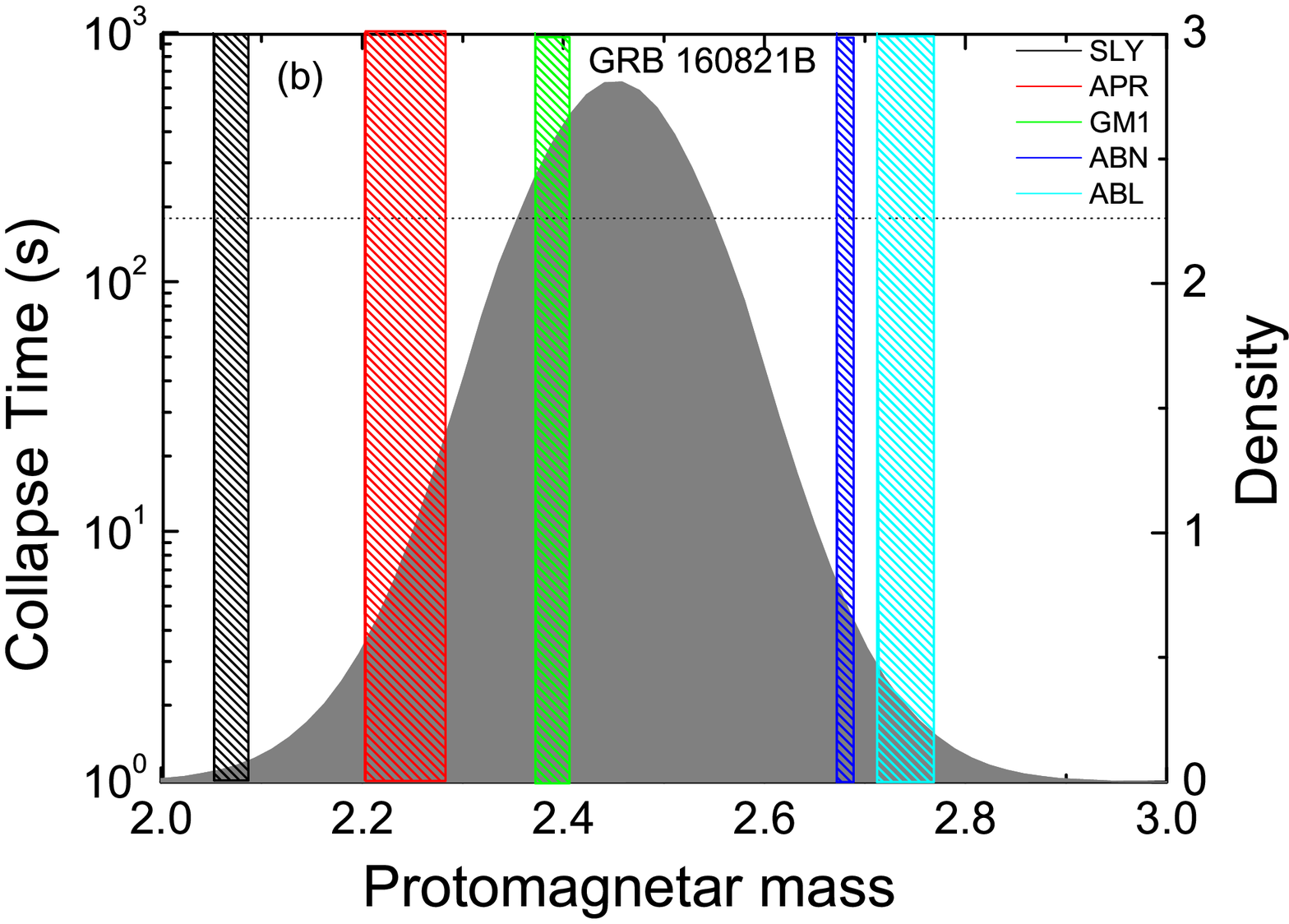}
\hfill\caption{(a): Inferred magnetar parameters, initial spin period $P_0$ vs.
surface polar cap magnetic field strength $B_p$ derived for GRB 160821B (red star),
compared with other short GRBs (grey triangle) and GRB 090515 (blue point). The
vertical solid line is the break-up spin period limit for a neutron star (Lattimer \&
Prakash 2004). (b): Collapse time as a function of the protomagnetar mass of GRB 160821B for
different EOS: SLy (black), APR (red), GM1 (green), AB-N (blue), and AB-L (cyan).
The horizontal dotted line is the observed collapse time.}
\end{figure}


\begin{figure}
\includegraphics[angle=0,scale=0.6]{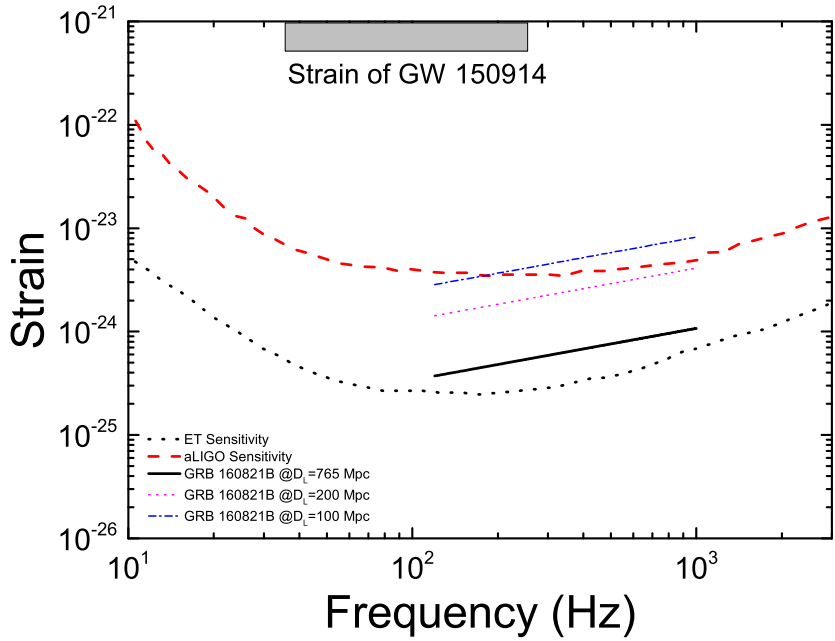}
\hfill\caption{Gravitational wave strain evolution with frequency for GRB 160821B,
at distance $D_{\rm L}=765~\rm Mpc$ (black solid line), $200~\rm Mpc$ (pink dot line),
$100~\rm Mpc (blue dash-dot line)$. The grey region is the strain of GW 150914 between
$35~\rm Hz$ and $250~\rm Hz$.
The black dotted line and red dashed line are the sensitivity limits for aLIGO and the Einstein
Telescope, respectively.}
\end{figure}


\end{document}